\documentclass[slac_one]{revtex4}
\usepackage{graphicx}
\usepackage{fancyhdr}
\usepackage{longtable} 
\usepackage{subfigure}
\begin{document}

\title{Vector Boson Fusion Higgs to Tau Tau Searches at the ATLAS Experiment} 

%

\author{K.J.C. Leney, on behalf of the ATLAS Collaboration}
\affiliation{University of Liverpool, United Kingdom}

\begin{abstract}
The search for a Higgs boson produced via Vector Boson Fusion and subsequently decaying to two tau leptons is discussed.  Significances for the di-lepton and lepton-hadron decay channels are presented, and the fully hadronic decay channel is shown to be feasible in terms of trigger, mass reconstruction and signal efficiency.  We consider performance issues for tau ID, missing $E_{T}$, forward jet identification, and central jet and $b$-jet vetoes, and outline several methods to estimate background contributions. 
\end{abstract}

\maketitle

\thispagestyle{fancy}


\section{INTRODUCTION} 
The second largest Higgs production process at the LHC will come from the fusion of vector bosons, radiated off initial state quarks (VBF)~\cite{PhysTDR}
.  The outgoing quarks receive a transverse deflection which gives rise to the distinctive signature of two high $p_{T}$ jets close to the beam-pipe in the ATLAS detector.  In addition, the lack of colour exchange between quarks leads to the suppression of QCD radiation between the two forward jets.  

The signal signature, and cuts to optimise event selection are discussed in section~\ref{section:signal}, followed by a summary of techniques to estimate contributions from the various background channels in section~\ref{section:background}.  Finally, in section~\ref{section:results} we present estimates for the expected signal significance and exclusion power.

The work presented here is based on the `Search for the Standard Model Higgs boson via Vector Boson Fusion production process in the di-tau channels with ATLAS' chapter from reference~\cite{CSCNote}.  Unless otherwise stated, $m_{H}$ is assumed to be 120 GeV.


\section{SIGNAL SELECTION}\label{section:signal}

The VBF Higgs $\to \tau \tau$ signal is characterised by two leptons ($e$, $\mu$) or $\tau$-jets in the central region of the detector, missing transverse energy from neutrinos and two (VBF) jets in the forward region of the detector.  Three final states are considered: Two leptons ($e$, $\mu$) in the final state ($\sigma$ = 0.036 pb); One leptonically decaying $\tau$, plus one hadronically decaying $\tau$ ($\sigma$ = 0.134 pb);  Two hadronically decaying $\tau$'s ($\sigma$ = 0.123 pb).  

\begin{figure*}[!ht]
\centering
\subfigure[]{
\includegraphics[height=45mm]{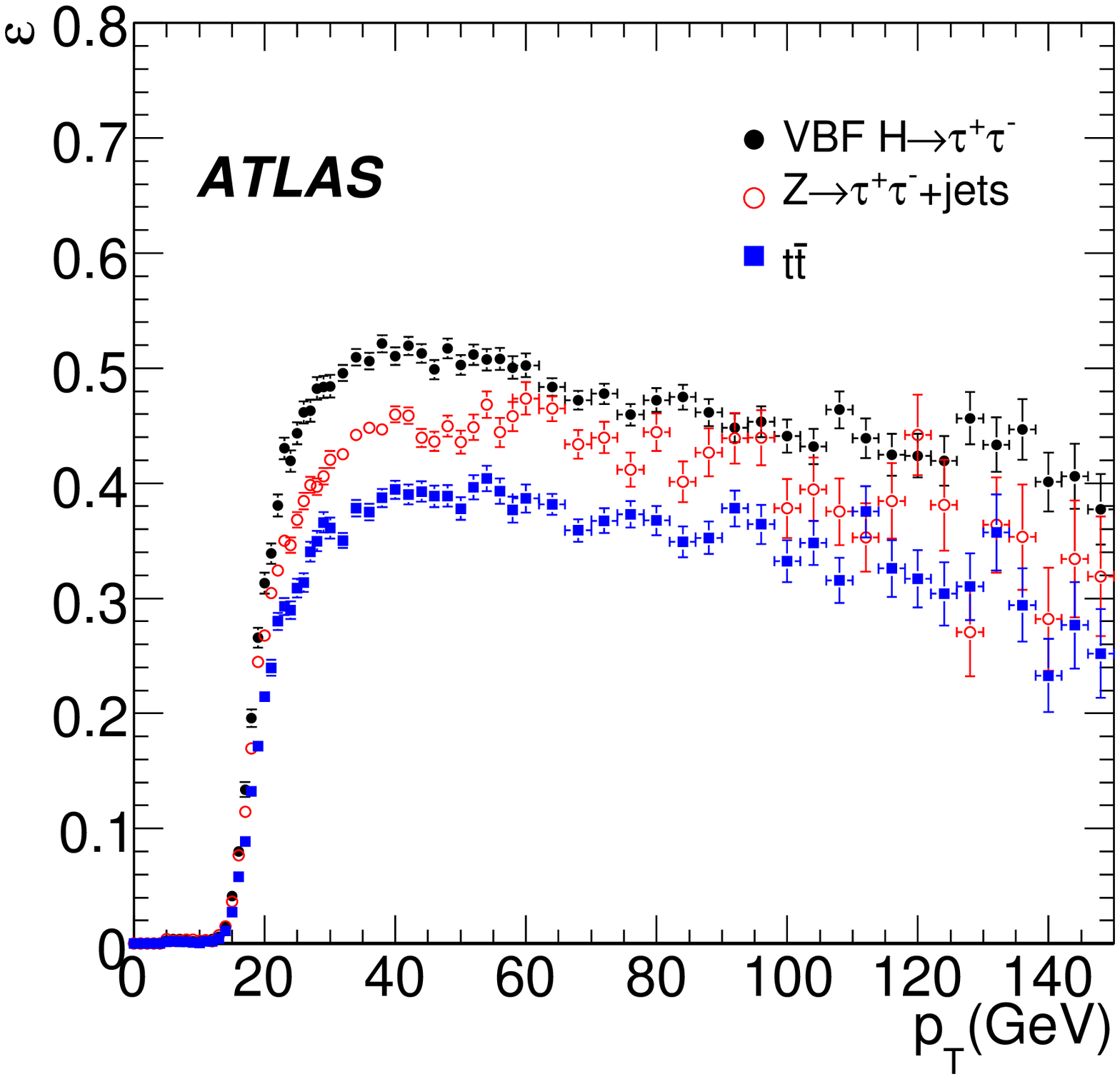}
}
\subfigure[] {
\includegraphics[height=45mm]{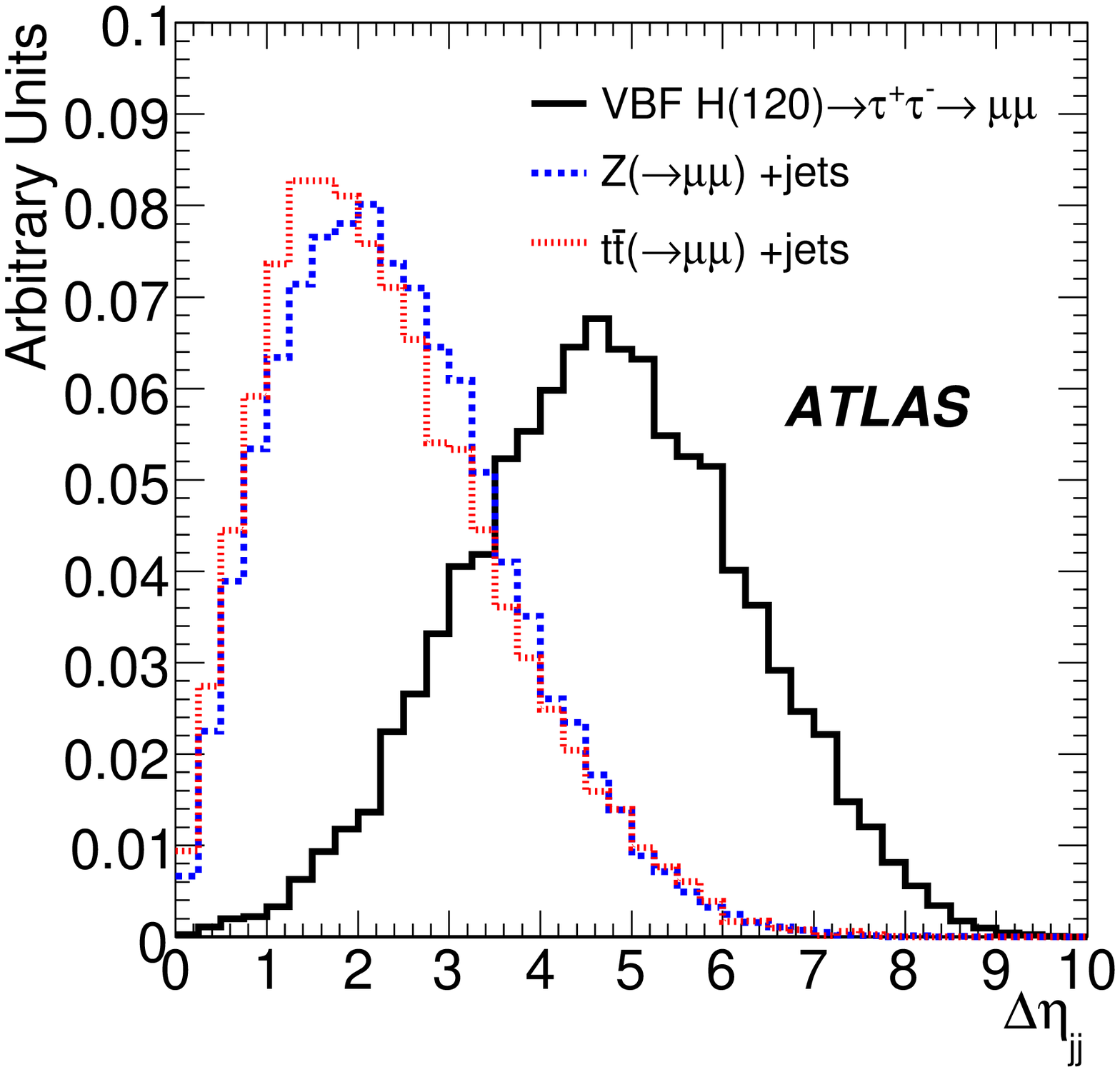}
}
\caption{Reconstruction and identification efficiency of the hadronic tau (a), and the pseudorapidity gap between the two forward jets (b).\label{fig:HadTauEff}} 
\end{figure*}

Approximately 65\% of tau leptons decay hadronically, the majority of these taking single prong decays\cite{PDGBook}.   A calorimeter cluster seeded algorithm is used for hadronic tau identification and reconstruction in this analysis.  This efficiency as a function of $p_{T}$ is shown, both for the $H\to\tau\tau$ signal, and the two main backgrounds: QCD $Z\to\tau\tau$ and $t\bar{t}$, in figure~\ref{fig:HadTauEff}.  Although ATLAS offers several possibilities for triggering that take advantage of the signal's complex final state, this analysis uses only the simplest, most robust triggers available~\cite{CSCNote}.  Hadronic tau and missing energy triggers are only used in the fully hadronic channel.  The other two channels simply require a trigger on either an electron with $p_{T} >$ 25 GeV, or a muon with $p_{T} >$ 20 GeV.

Significant missing energy is always present in an event due to the neutrinos from tau decays.  If one makes the approximation that the tau decay products are collinear with the tau in the laboratory frame, it is possible to express the tau momentum in terms of the momentum fraction carried by the tau daughter.  It is thus possible to reconstruct the invariant mass of the tau pair.  Since $M_{H}/2 \gg M_{\tau}$, the taus are heavily boosted and the approximation works well.  A complete discussion can be found in~\cite{CSCNote}.

The VBF production process generally leads to a large separation in $\eta$ of the two highest $p_{T}$ jets in the event, and the value $\Delta \eta_{jet-jet}$ provides an excellent discriminating variable.  Discarding any event where there is an additional high $p_{T}$ jet in the central region is particularly effective at removing QCD backgrounds.  

The largest background contribution to the di-lepton channel comes from $t\bar{t}$ decays.  $b$-jets from the top quark decay can be misidentified as the VBF jets in the signal.  Including a $b$-jet veto into the analysis reduces the $t\bar{t}$ background by a factor of 2-3 with little effect on the signal.  


\section{BACKGROUND ESTIMATION}\label{section:background}

The main processes contributing to background come from $Z\to ee, \mu\mu, \tau\tau + N$ partons (QCD and Electroweak), $W \to e\nu, \mu\nu + N$ partons, $WW, WZ, ZZ$ and $t\bar{t} \to WbWb$.  Of these, the main contributions come from the $t\bar{t}$ and $Z\to\tau\tau$ processes.  Various techniques have been developed to provide an estimate of background contributions.  These include, but are not limited to, the methods outlined below.

\subsection{Momentum Scaling Method}
\begin{figure*}[!ht]
\centering
\subfigure[]{
\includegraphics[height=45mm]{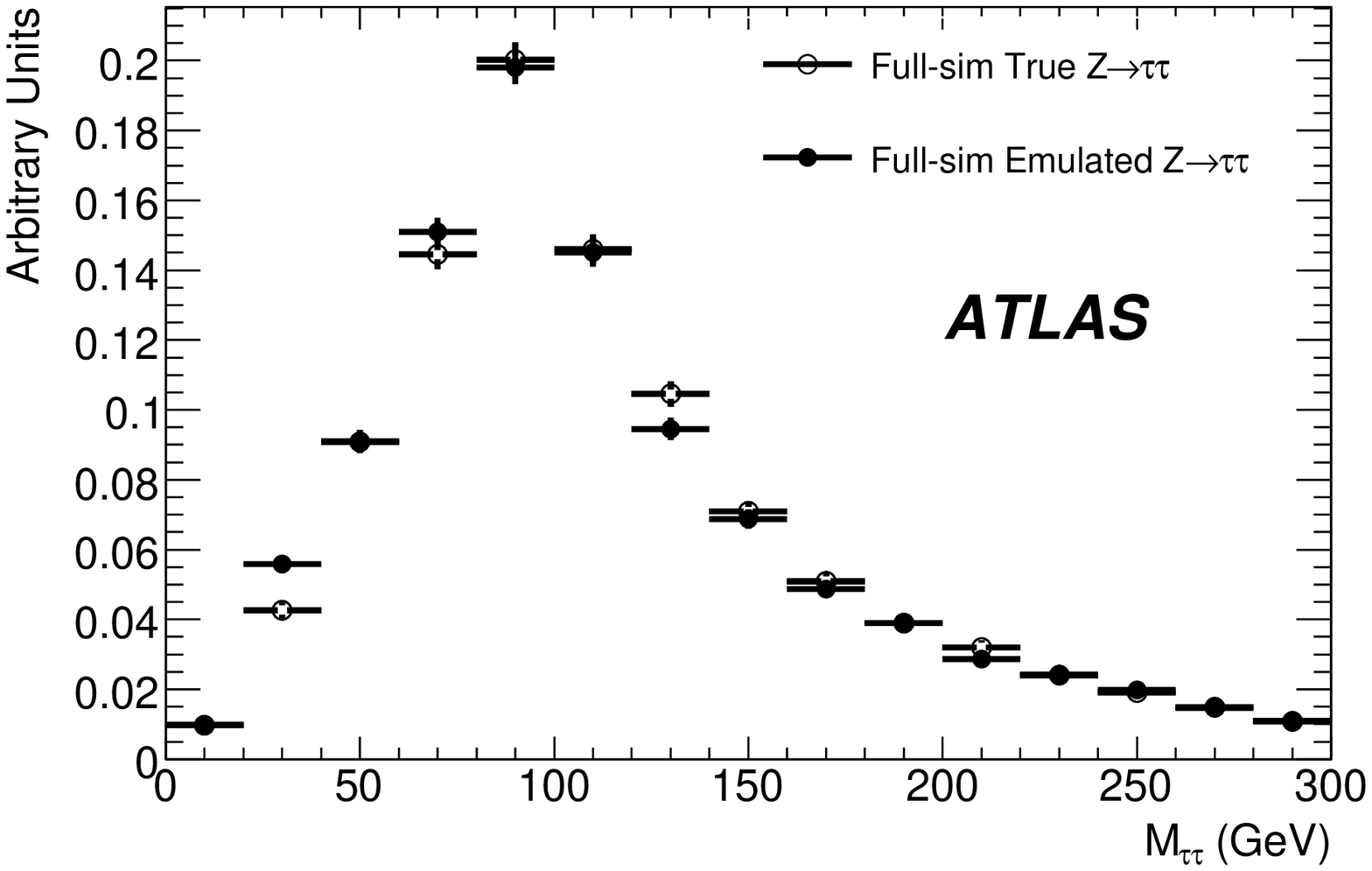}
}
\subfigure[] {
\includegraphics[height=45mm]{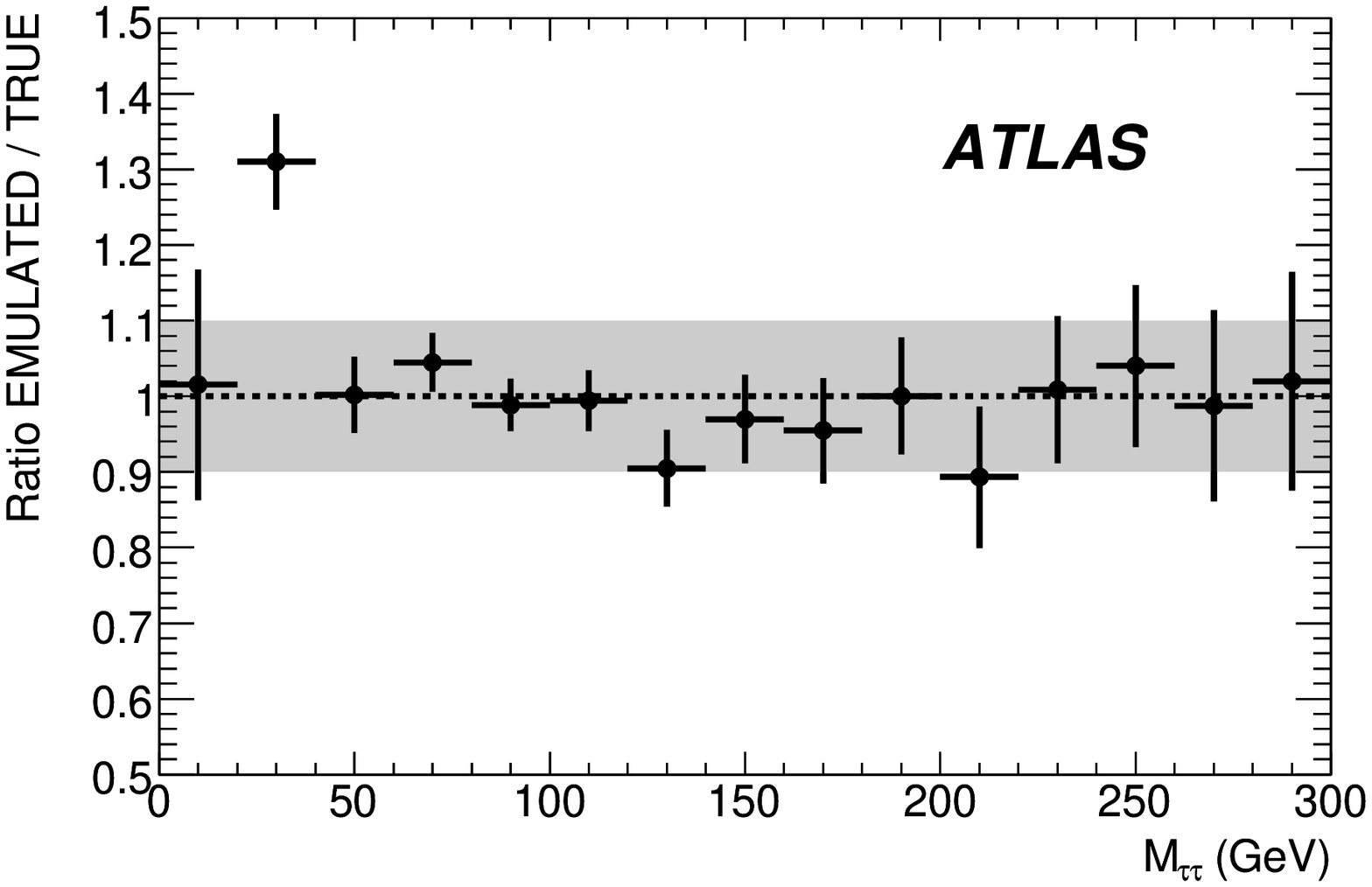}
}
\caption{Reconstructed invariant mass distribution (a) and bin-by-bin ratio generated from the true and emulated $Z\to\tau\tau\to lh$ events (b).} \label{fig:Trevor}
\end{figure*}

It will be relatively easy to obtain a clean sample of  $Z\to\mu\mu$ events.  The muon's measured momentum vector is then used to seed an equivalent Monte Carlo tau.  Once the two muons have been decayed as taus, they are passed to the ATLAS detector simulation and reconstruction software.  This provides data-driven control samples for all the decay channels.  Several comparisons of the $Z\to\tau\tau$ sample emulated from $Z\to\mu\mu$ events and the true $Z\to\tau\tau$ sample were made and show excellent agreement.  The reconstructed invariant mass distribution, and the bin-by-bin ratio of true to emulated $Z\to\tau\tau$ events are shown in figure~\ref{fig:Trevor}.  The grey band represents $\pm$10\% around a ratio of 1.

\subsection{Track Multiplicity}
The $t\bar{t}$ background process includes contributions both from real taus, and from jets which fake hadronic tau decays ($\sim$50\% in the lepton-hadron channel).  Real taus typically have one or three tracks, with some spread due to tracking efficiency, or the presence of spurious tracks. Electrons usually have a single track, while jets have a broad multiplicity spectrum.  By comparing the inclusive distribution with those obtained from just the electrons and taus (which can be obtained from data), one can determine the overall QCD contribution.

\subsection{Cut Factorisation Method}
Often, not enough Monte Carlo statistics are available to calculate background suppression directly.  In these cases, the analysis cuts are divided into 3 categories: those which are related to taus, those concerned with jets, and finally, those cuts which have a strong correlation to both.  Each set of cuts can then be applied in turn, and factorised to obtain an overall rejection efficiency.

\subsection{Summary of Background Predictions}
Table~\ref{table:crossSections} summarises the predicted final cross-sections for signal and background processes in each channel.  Statistical uncertainties arising from the limited size of the Monte Carlo samples are also given.

\begin{table}[!ht]
\begin{center}
\begin{tabular}{|l||c|c|c|c|c|c|c|c|c|}\hline
  	&$Signal$  &\multicolumn{2}{c|}{$Z\rightarrow\tau\tau+jets$}  &$t\bar{t}$  &$Z\rightarrow ll$  &$W\rightarrow l\nu$  &$Di-Boson$  &$W\rightarrow \tau\nu$  &$QCD$ 	\\
	&	&$QCD$		&$EW$		&		&$+jets$		&$+jets$		&$(WW,WZ,ZZ)$	&$+jets$		&$Di-Jet$ \\\hline\hline
$ll$	&0.45(1)		&0.23(1)		&0.04(1)		&0.10(2)$^{*}$	&0.058(3)$^{*}$	&0.01(1)$^{*}$	&0.002(1)	& - 		& -\\ 
$lh$	&0.61(2)		&0.11(2)		&0.04(1)		&0.012(5)$^{*}$	&0.008(1)$^{*}$	&0.020(6)$^{*}$	&0.001(1)	& - 		& - \\
$hh$	&0.34(2)		&0.08(3)$^{*}$	&0.003(1)	&0.03(3)$^{*}$	& -		& - 		& - 		&0.1(1)$^{*}$	&$1(1)^{*}$\\\hline

\end{tabular}
\caption{Initial and final cross-sections (in fb) for signal and background processes after applying all selection cuts, including a mass window around a Higgs mass of 120 GeV.  An asterisk is used to indicate cross-sections estimated by the cut factorisation method.}\label{table:crossSections}
\end{center}
\end{table}

An analysis in the fully hadronic channel appears feasible in terms of trigger, mass reconstruction and signal efficiency (roughly comparable to di-lepton and lepton-hadron channels).  However, although the effects of $Z+jets$, $W+jets$, di-boson, and $t\bar{t}$ backgrounds are understood, the pure QCD background will be dominant, and this can only be reliably estimated with data.  For this reason, we do not report an estimate of the sensitivity for this channel here.


\section{RESULTS}\label{section:results}

\begin{figure*}[!ht]
\centering
\subfigure[Di-Lepton Channel]{
\includegraphics[height=45mm]{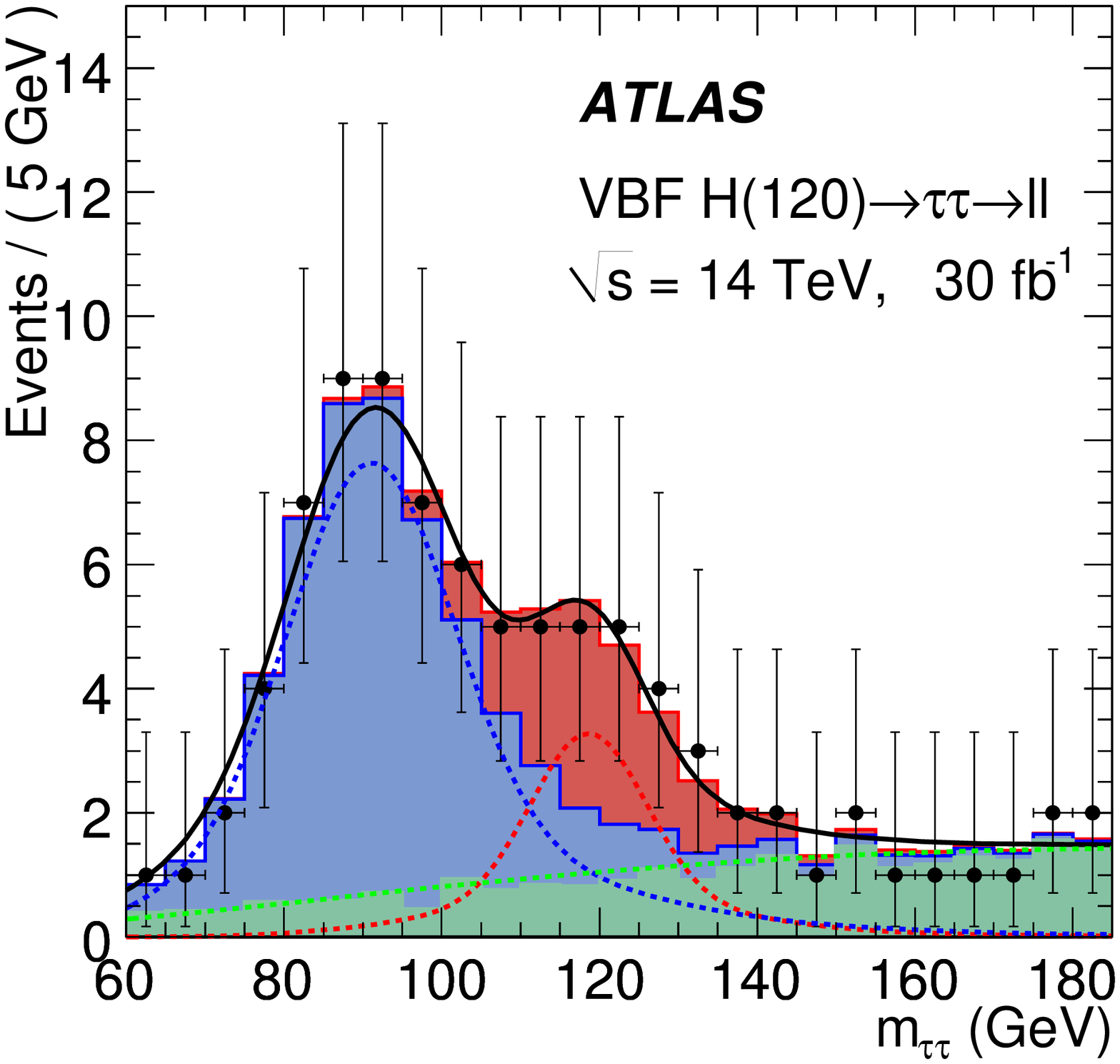}
}
\subfigure[Lepton-Hadron Channel] {
\includegraphics[height=45mm]{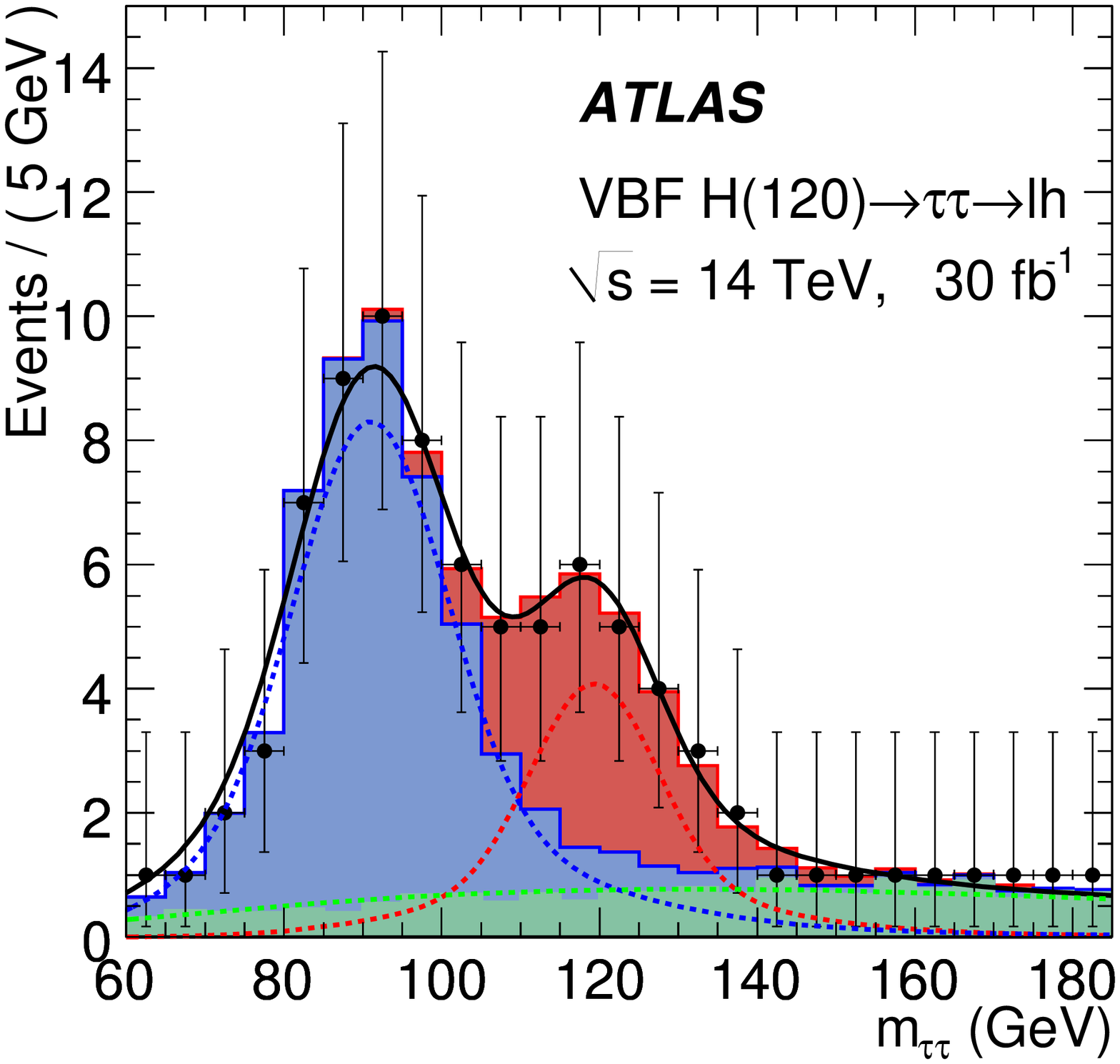}
}
\caption{Example fits to a data sample for the di-lepton (a) and lepton-hadron(b) channels with 30 fb$^{-1}$ of data.}\label{fig:MassReco}
\end{figure*}

The invariant mass of the tau pair is reconstructed using the collinear approximation.  Example fits to signal-plus-background models for the di-lepton and lepton-hadron models are shown in figure~\ref{fig:MassReco}

These results do not include the effects of pile-up;  Extra proton-proton interactions will cause additional hadronic activity which can cause events to fail the central jet veto.  Under pile-up conditions, the probability of a signal event surviving this cut falls from $\sim$88\% to $\sim$75\%.  The presence of pile-up also degrades the missing $E_{T}$ resolution and $\tau$ identification performance which leads to a drop in the mass resolution from $\sim$9.5 to $\sim$11.5 GeV for $m_{H}$ = 120 GeV.  

\begin{figure*}[!ht]
\centering
\subfigure[]{
\includegraphics[height=45mm]{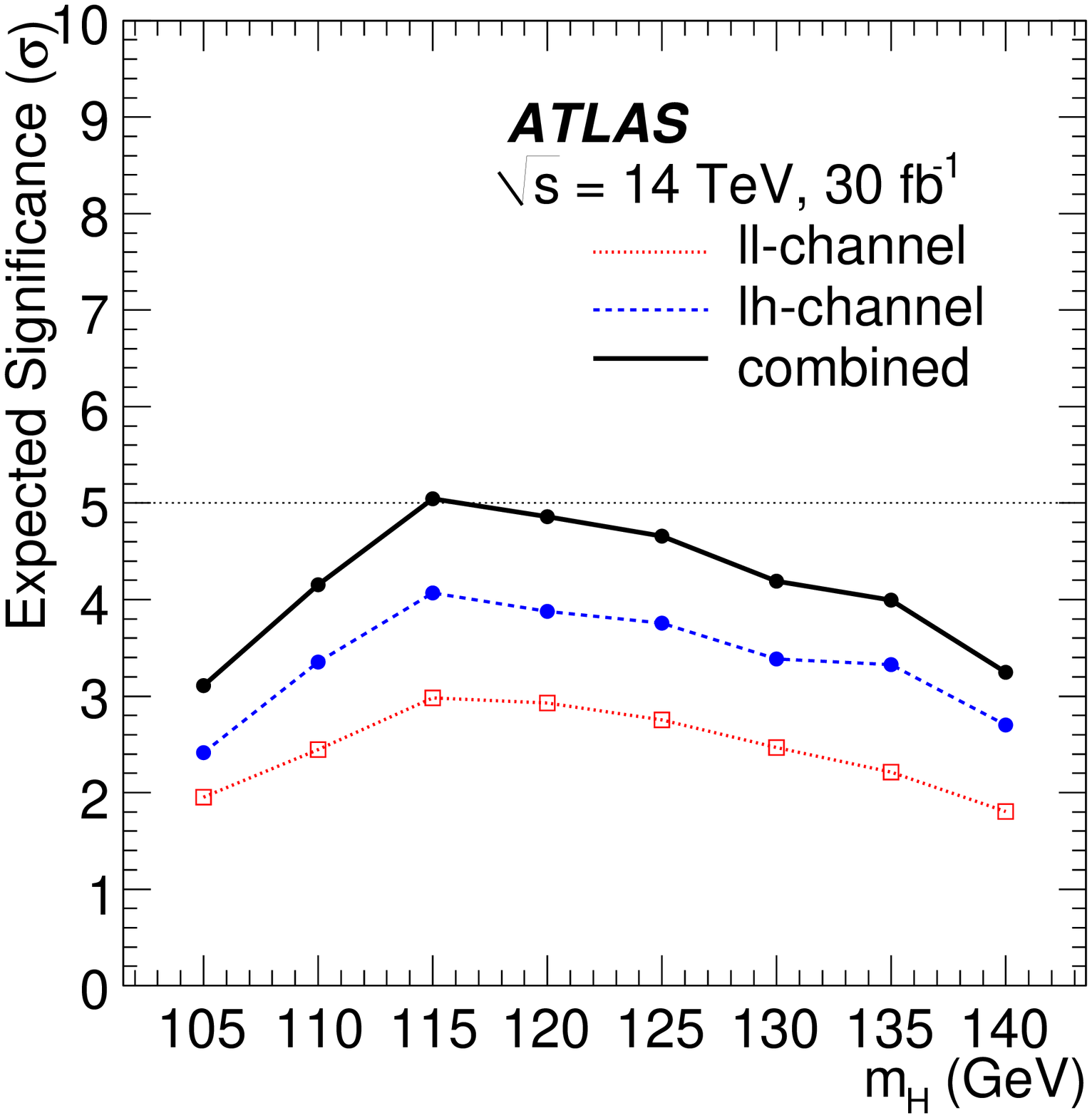}
}
\subfigure[] {
\includegraphics[height=45mm]{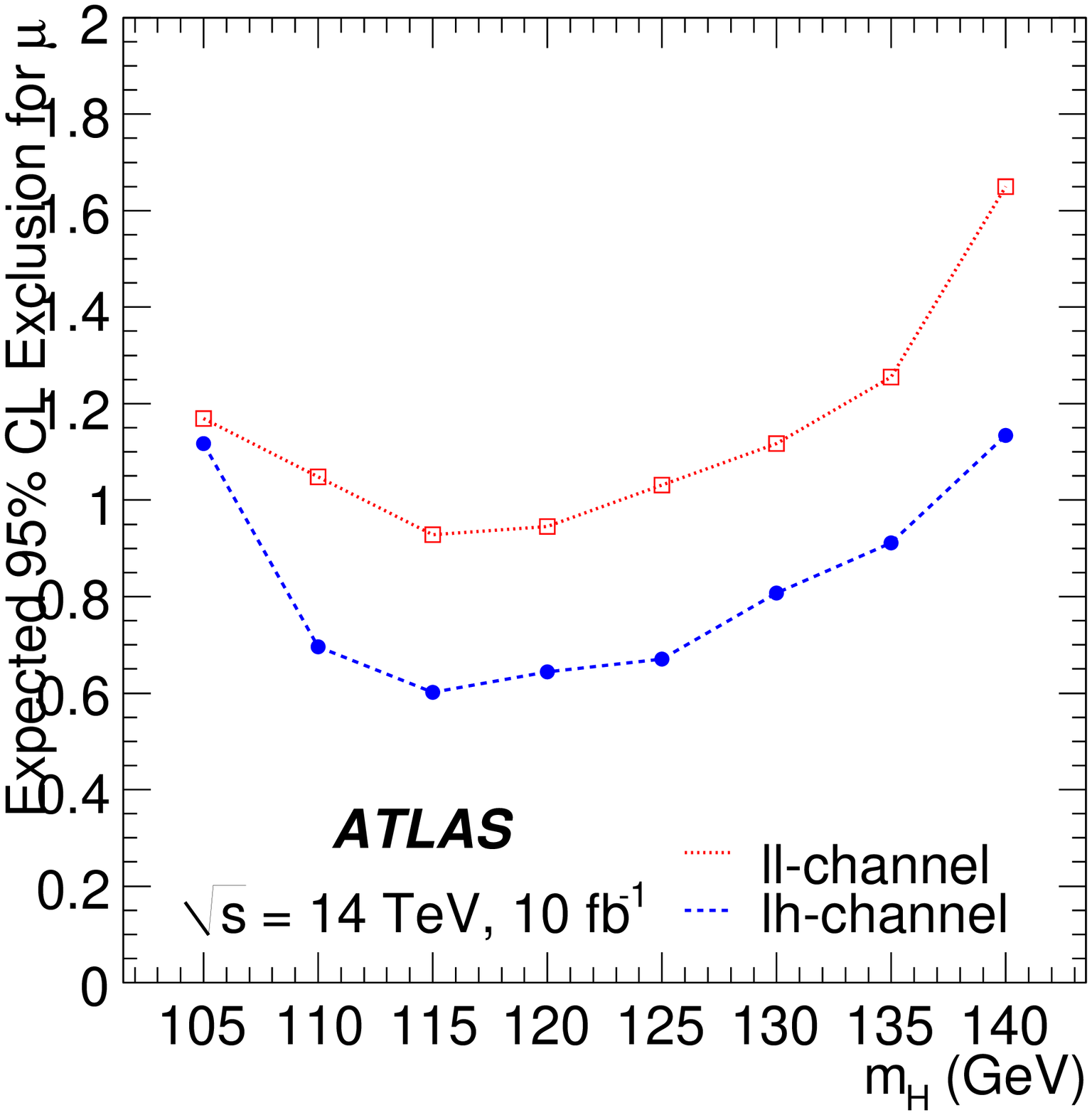}
}
\caption{Expected signal significance (a), and 95\% exclusion of the signal rate in units of the Standard Model expectation, $\mu$, (b), for a range of Higgs boson masses.\label{fig:SigEx}}
\end{figure*}

The expected significance and the exclusion power based on the uncertainty in the signal efficiency, across a range of masses for the di-lepton and lepton-hadron channels are shown in figure~\ref{fig:SigEx}.  The data-driven background estimation methods described in section~\ref{section:background} have been developed so that uncertainty in the background shape and normalisation are included directly into the significance calculation.  A full discussion of how the significance is estimated and the treatment of systematic uncertainties can be found in reference~\cite{CSCNote}.  Results (without pile-up) indicate that a significance of $\sim 5 \sigma$ can be achieved for a Higgs boson with a mass in the range 115-125 GeV after collecting 30 fb$^{-1}$ of data and combining the di-lepton and lepton-hadron channels.  This may be improved by the use of multivariate techniques or by further optimisation of the analysis cuts.

\end{document}